\documentclass{rspublic}

\usepackage{natbib}
\usepackage{graphicx}

\begin{document}

\title[Brain architecture for natural computation]{Brain architecture: A design for natural computation}

\author[M. Kaiser]{Marcus Kaiser\thanks{Author for correspondence (M.Kaiser@ncl.ac.uk).}}
%\TandT\
\affiliation{School of Computing Science, Newcastle University, Claremont Tower, Newcastle upon Tyne NE1 7RU, UK \\ Institute of Neuroscience, Newcastle University, Framlington Place, Newcastle upon Tyne NE2 4HH, UK}

\label{firstpage}

\maketitle

\begin{abstract}{neural networks, computational neuroanatomy, network science, spatial graph, robustness, recovery}
Fifty years ago, John von Neumann compared the architecture of the brain with that of computers that he invented and which is still in use today. In those days, the organisation of computers was based on concepts of brain organisation. Here, we give an update on current results on the global organisation of neural systems. For neural systems, we outline how the spatial and topological architecture of neuronal and cortical networks facilitates robustness against failures, fast processing, and balanced network activation. Finally, we discuss mechanisms of self-organization for such architectures. After all, the organization of the brain might again inspire computer architecture.
\end{abstract}

\section{Introduction}
The relation between the computer and the brain has always been of interest to scientists and the public alike. From the notion of 'thinking machines' and 'artificial intelligence' to applying concepts of neuroscience such as neural networks to solve problems in computer science. Also the earliest computers, using the von Neumann architecture still in use today, used memory and a central processing unit based on concepts of brain architecture \citep{vonNeumann1958}. Also, models of artificial neural networks were inspired by the function of individual neurons as integrators of incoming signals. Detailed models of neural processing, however, are often limited to single tasks (e.g., pattern recognition) and one modality (e.g., only visual information). In addition, artificial neural networks starting with Perceptrons \citep{Rosenblatt1959} are designed as a general purpose architecture whereas the architecture of natural neural systems shows a high specialization according to different tasks and functions. Global models, on the other hand, often deal with functional circuits (e.g. movement planning) without a direct link to the local structure of the neural network. Therefore, much of the complexity of neural processing in terms of combining local and global levels as well as integrating information from different domains is largely missing from current models.

About 50 years ago, John von Neumann---inventor of the current computer architecture---thought about where computers and the brain are the same or where they differ \citep{vonNeumann1958}. After 50 years of technological progress, how do the benchmark characteristics differ? The human brain consists of 10$^{10}$ neurons or processing units. The Internet, being the largest computer network, has only millions of processing units. However, the extension of the Internet to mobile services (pervasive computing) could lead to billions of processing nodes in the future. The human memory can be estimated from adjustable synaptic weights of connections between neurons. However, these 10$^{14}$ synapses/weights are only a first approximation of the hard-wired information storage as the position of synapses, both absolute on the target neuron and relative to other synapses influences signal integration. Computer memories have reached this level with some systems, such as the machines that store web information at Google, storing several petabytes (1 petabyte=10$^{15}$ bytes, see http://en.wikipedia.org/wiki/Petabyte). However, computer systems are still far-away from processing complex information like the human brain does. In spite of processing units or memory, the main difference between computers and brains is their hardware architecture---how they are wired up.

In this article, we present recent results on the topology (architecture) of complex brain networks. These results are not about standard (artificial) neural networks that deal with one single task, e.g. face recognition. Rather, we look at the high-level organization of the brain including modules for different tasks and different sensory modalities (e.g., sound, vision, touch). Nonetheless, similar organization \citep{Buzsaki2004} and processing \citep{Dyhrfjeld-Johnsen2007} has been found at the local level of connectivity within modules.

\section{Cortical network organization}
\subsection{Cluster organization}
\begin{figure}[htbp]
	\centering
		\includegraphics[width=15cm]{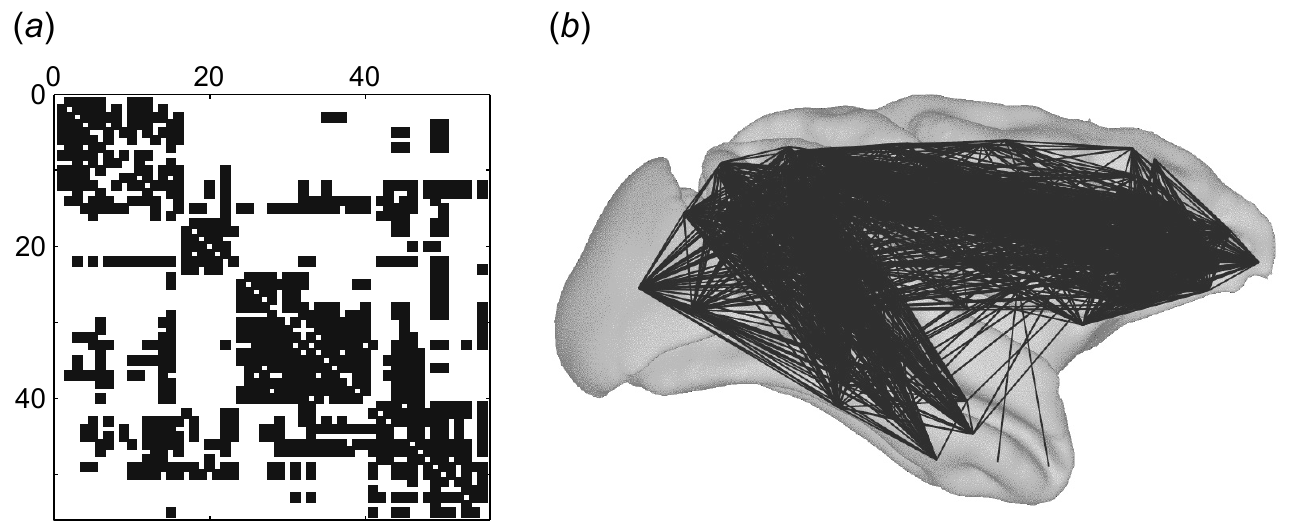}
	\caption{({\it a}) Adjacency Matrix of the cat connectivity network (55 nodes; 891 directed edges). Dots represent 'ones' and white spaces the 'zero' entries of the adjacency matrix. ({\it b}) Macaque cortex (95 nodes; 2,402 directed edges).}
	\label{fig1cortex}
\end{figure}

Cortical areas are brain modules which are defined by structural (microscopic) architecture. Observing the thickness and cell types of the cortical layers, several cortical areas can be distinguished \citep{Brodmann1909}. Furthermore, areas also show a functional specialization. Within one area further sub-units (cortical columns) exist, however, these units will not be covered in this review as there is not enough information about their connectivity. Using neuro-anatomical techniques, it can be tested which areas are connected, that means that projections in one or both directions between the areas do exist. If a fiber projection between two areas is found, the value '1' is entered in the adjacency matrix; the value '0' defines absent connections or cases where the existence of connections was not tested (figure \ref{fig1cortex}{\it a}).

Contrary to popular belief, cortical networks are not completely connected, i.e. {\it not} 'everything is connected to everything else': Only about 30\% of all possible connections (arcs) between areas do exist.  Instead, highly connected sets of nodes ({\it clusters}) are found that correspond to functional differentiation of areas. For example, clusters corresponding to visual, auditory, somatosensory and fronto-limbic processing were found in the cat cortical connectivity network \citep{Hilgetag2004}. Furthermore, about 20\% of the connections are unidirectional \citep{Felleman1991}, i.e. a direct projection from area A to area B but not vice versa exists. Although some of these connections might be bidirectional as the reverse direction was not tested, there were several cases where it was confirmed that projections were unidirectional. Therefore, measures that worked for directed graphs were used.

Until now, there is not enough information about connectivity in the human brain that would allow network analysis \citep{Crick1993}. However, several new non-invasive methods including diffusion tensor imaging \citep{Tuch2005} and resting state networks \citep{Achard2006} are under development and might help to define human connectivity in the future.  At the moment, however, we are bound to analyze known connectivity in the cat and the macaque (rhesus monkey,  Fig. \ref{fig1cortex}{\it b}) cortical networks \citep[see also][]{Passingham2002,Sporns2004}. Both networks exhibit clusters, i.e. areas belonging to a cluster have many existing connections between them but there are few connections to areas of different clusters \citep{Young1993,Scannell1995}. These clusters are also functional and spatial units. Two connected areas tend to be spatially adjacent on the cortical surface and tend to have a similar function (e.g., both taking part in visual processing). Whereas there is a preference for short-length connections to spatially neighboring areas for the macaque, about 10\% of the connections cover a long-distance ($\geq 40$ mm) -- sometimes close to the maximum possible distance (69 mm) between two areas of one hemisphere \citep{Kaiser2004c}.

Cortical networks show maximal structural and dynamic complexity which is thought to be necessary for encoding a maximum number of functional states and might arise as a response to rich sensory environments \citep{Sporns2000}. Using methods and concepts of network analysis \citep{Albert2002}, we discuss small-world and scale-free properties as well as motifs and spatial network organization of cortical networks.

\subsection*{Small-world properties}
Many complex networks exhibit properties of small-world networks \citep{Watts1998}. In these networks neighbors are better connected than in comparable Erd\"os-R\'{e}nyi random networks \citep{Erdoes1960} (called random networks throughout the text) whereas the average path length remains as low as in random networks. Formally, the average shortest path (ASP, similar, though not identical, to characteristic path length $\ell$ \citep{Watts1999}) of a network with $N$ nodes is the average number of edges that has to be crossed on the shortest path from any one node to another:
\begin{equation}\label{asp}
ASP = \frac{1}{N (N-1)} \sum_{i, j} d(i,j)   \ \ \ \ \ with \ i\ne j,
\end{equation}
where $d(i,j)$ is the length of the shortest path between nodes $i$ and $j$.

The neighborhood connectivity is usually measured by the clustering coefficient. The clustering coefficient of one node $v$ with $k_v$ neighbors is
\begin{equation}\label{clustercoef}
C_v = \frac{|E(\Gamma_v)|}{{k_v \choose 2}},
\end{equation}
where $|E(\Gamma_v)|$ is the number of edges in the neighborhood of $v$ and ${k_v \choose 2}$ is the number of possible edges \citep{Watts1999}. In the following analysis, we use the term clustering coefficient as the average clustering coefficient for all nodes of a network.

Small-world properties were found on different organizational levels of neural networks: from the tiny nematode {\it C. elegans} with about 300 neurons \citep{Watts1998} to cortical networks of the cat and the macaque \citep{Hilgetag2000b,Hilgetag2004}. Whereas the clustering coefficient for the macaque is 49\% (16\% in random networks), the ASP is comparably low with 2.2 (2.0 in random networks). That is, on average only one or two intermediate areas are on the shortest path between two areas. Note that a high clustering coefficient does not necessarily correlate with the existence of multiple clusters. Indeed, the standard model for generating small-world networks by rewiring regular networks \citep{Watts1998} does not lead to multiple clusters.

\section{Robustness and recovery}
Compared to technical networks (power grids or communication networks), the brain is remarkably robust towards damage. On the local level, Parkinson's disease in humans only becomes apparent after more than half of the cells in the responsible brain region are eliminated \citep{Damier1999}. On the global level, the loss of the whole primary visual cortex (areas 17, 18 and 19) in kittens can be compensated by another region, the postero-medial supra-sylvian area (PMLS) \citep{Spear1988}. On the other hand, the removal of a small number of nodes or edges of the network can lead to a breakdown of functional processing. As functional deficits are not related to the number or size of removed connections or brain tissue, it might be the role within the network that makes some elements more critical than others. Identifying these critical components has applications in neurosurgery where important parts of the brain should remain intact even after the removal of a brain tumour and its surrounding tissue.

\subsection*{Critical connections in neural systems}
\begin{figure}[htbp]
	\centering
		\includegraphics[width=\textwidth]{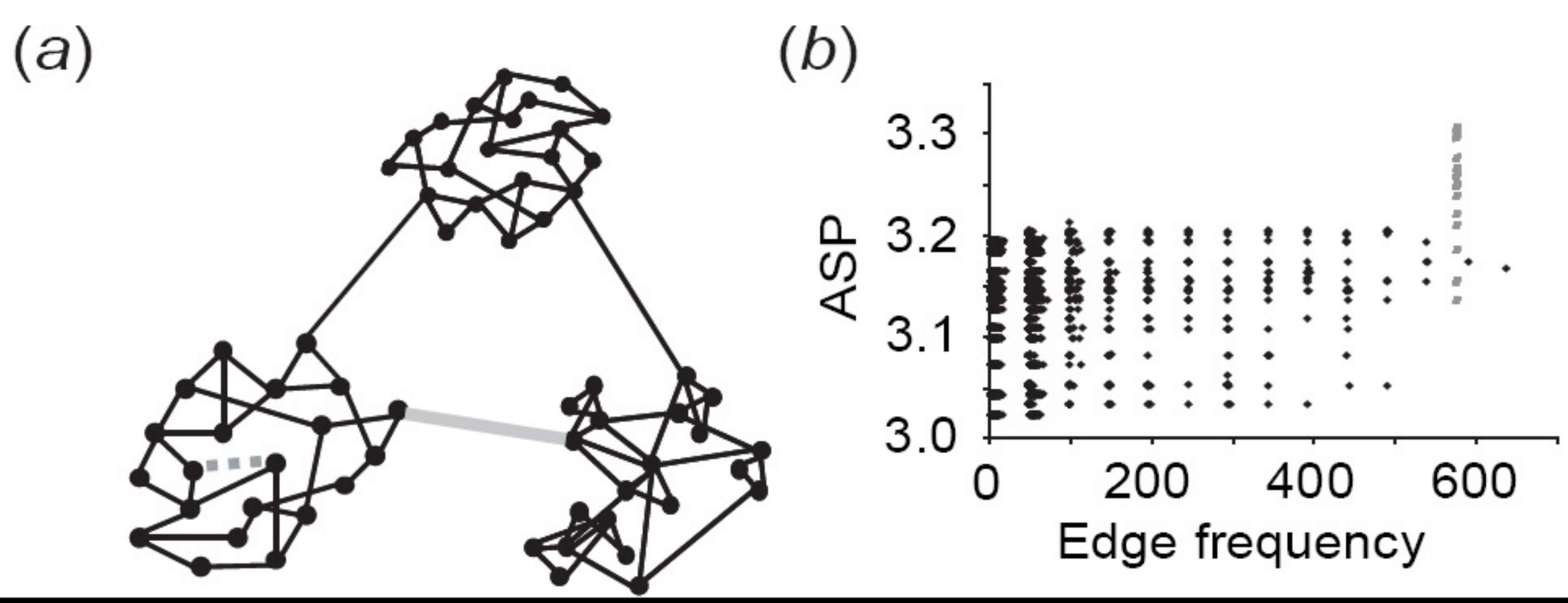}
	\caption{({\it a}) Schematic drawing of a network with three clusters showing examples for an intra- (gray dashed line) and inter-cluster (gray solid line)  connection. ({\it b}) Edge frequency of the eliminated edge vs. ASP after edge removal (20 generated networks with three clusters, defined inter-cluster connections and random connectivity within clusters; inter-cluster connections: light-gray; connections within a cluster: black).}
	\label{fig2sw}
\end{figure}

It was found that the robustness towards edge removal is linked to the high neighborhood connectivity and the existence of multiple clusters \citep{Kaiser2004d}. For connections within clusters, many alternative pathways of comparable length do exist once one edge is removed from the cluster (figure \ref{fig2sw}{\it a}). For edges between clusters, however, alternative pathways of comparable length are unavailable and removal of such edges should have a larger effect on the network. The damage to the macaque network was measured as the increase of the ASP after single edge removal. Among several measures, edge frequency (approximate measure of edge betweenness) of an edge was the best predictor of the damage after edge elimination (linear correlation r=0.8 for macaque). The edge frequency of an edge counts the number of shortest paths in which the edge is included.

Furthermore, examining comparable benchmark networks with three clusters, edges with high edge frequency are the ones between clusters. In addition, removal of these edges causes the largest damage as increase in ASP (figure \ref{fig2sw}{\it b}). Therefore, inter-cluster connections are critical for the network. Concerning random loss of fiber connections, however, in most cases one of the many connections within a cluster will be damaged with little effect on the network. The chances of eliminating the fewer inter-cluster connections are lower. Therefore, the network is robust to random removal of an edge \citep{Kaiser2004d}.

\subsection*{Node removal behaviour similar to that of scale-free networks}
In addition to high neighborhood clustering, many real-world networks have properties of scale-free networks \citep{Barabasi1999}. In such networks, the probability for a node possessing $k$ edges is  $P(k)\propto k^{-\gamma}$. Therefore, the degree distribution---where the degree of a node is the number of its connections---follows a power-law.  This often results in highly connected nodes that would be unlikely to occur in random networks.  Technical networks such as the world wide web of links between web pages \citep{Huberman1999} and the Internet \citep{Faloutsos1999} at the level of connections between domains/autonomous systems. Do cortical networks, as natural communication networks, share similar features?

In cortical networks, some structures (e.g. evolutionary older structures like the Amygdala) are highly connected. Unfortunately, the degree distribution can not be tested directly as less than 100 nodes are available in the cat and macaque cortical networks. However, using the node elimination pattern as an indirect measure, cortical networks were found to be similar to scale-free benchmark networks \citep{Kaiser2007EJN}.

\begin{figure}[htbp]
	\centering
		\includegraphics{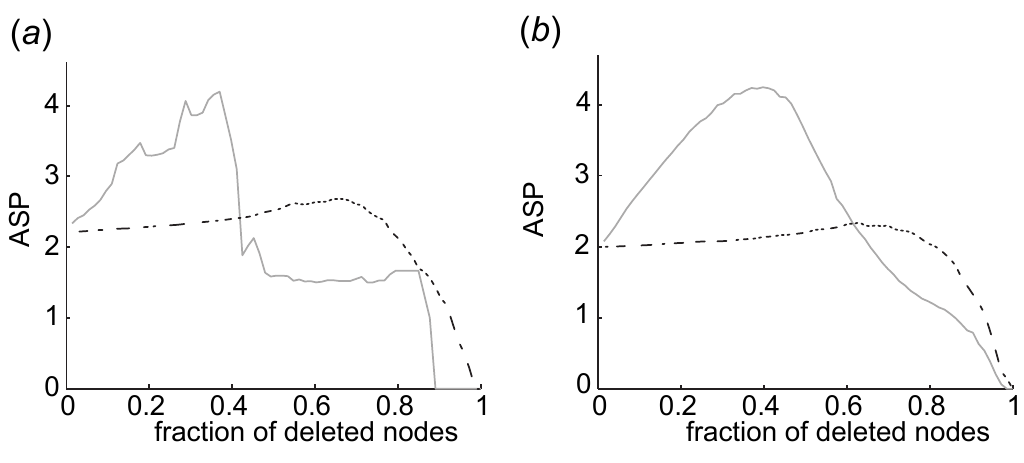}
	\caption{Average shortest path (ASP) after either random (dashed line) or targeted (gray solid line) subsequent node removal. ({\it a}) Macaque cortical network (73 nodes, 835 directed edges). ({\it b}) Scale-free benchmark network with the same number of nodes and edges (lines represent the average values over 50 generated networks and 50 runs each in the case of random node removal).}
	\label{fig3sf}
\end{figure}

In that approach, we tested the effect on the ASP of the macaque cortical network after subsequently eliminating nodes from the network until all nodes were removed \citep{Barabasi2000a}.  For random elimination, the increase in ASP was slow and reached a peak for a high fraction of deleted nodes before shrinking due to network fragmentation (figure \ref{fig3sf}{\it a}). When taking out nodes in a targeted way ranked by their connectivity (deleting the most highly connected nodes first), however, increase in ASP was steep and a peak was reached at a fraction of about 35\%. The curves for random and targeted node removal were similar for the benchmark scale-free networks (figure \ref{fig3sf}{\it b}) but not for generated random or small-world \citep{Watts1998} networks \citep{Kaiser2007EJN}. Therefore, cortical as well as scale-free benchmark systems are robust to random node elimination but show a larger increase in ASP after removing highly connected nodes. Again, as for the edges, only few nodes are highly connected and therefore critical so that the probability to select them randomly is low.

\section{Processing}
\subsection*{Wiring constraints for processing}
For microchips, increasing the length of electric wires increases the energy loss through heat dissipation. Inspired by these ideas, it was suggested that neural systems should be optimized to reduce wiring costs as well \citep{Cherniak1994}. In the brain, energy is consumed for establishing fibre tracts between areas and for propagating action potentials over these fibres. Thus, the total length of all wires should be kept as short as possible. This has led to the idea of optimal component placement in that modules are arranged in a way so that every rearrangement of modules would lead to an increase in total wiring length.

It has been proposed for several neural systems---including the {\it C. elegans} neural network and subsets of cortical networks---that components are indeed optimally placed \citep{Cherniak1994}. This means that all node position permutations of the network---while connections are unchanged---results in higher total connection length. Therefore, the placement of nodes is optimized to minimize the {\it total} wiring length. However, using larger data sets than used in the original study, we found that a reduction in wiring length by swapping the position of network nodes was possible.

\begin{figure}[htbp]
	\centering
		\includegraphics[width=\textwidth]{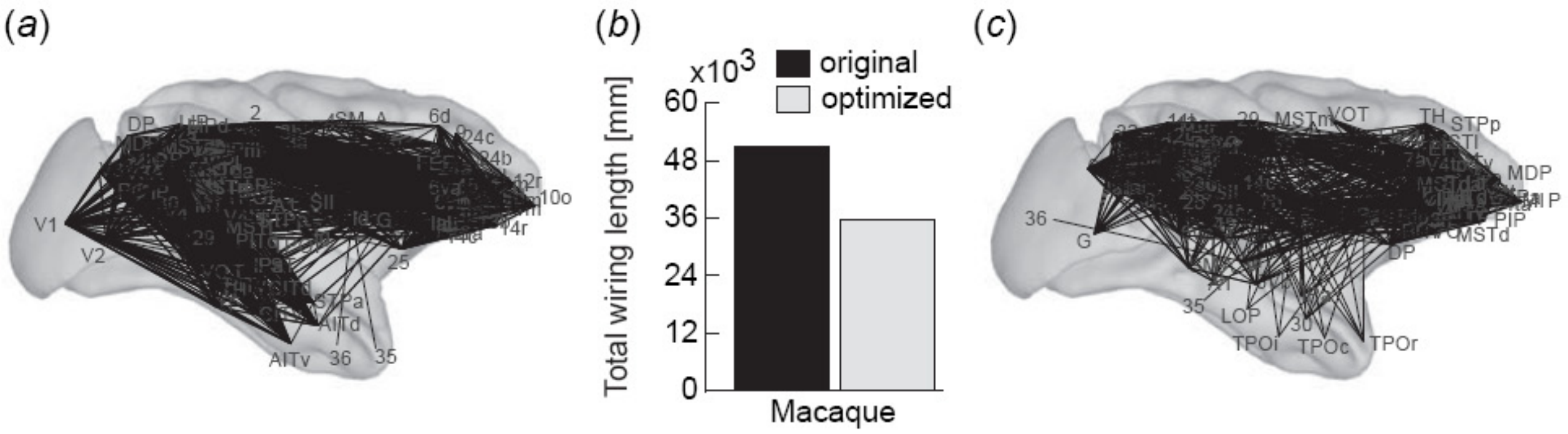}
	\caption{({\it a}) Original placement of cortical areas. ({\it b}) Wiring length optimization leads to a reduction in total wiring length by 32\% of the original length. ({\it c}) Placement after optimization for total wiring length.}
	\label{fig:wiring}
\end{figure}

For the macaque, we analyzed wiring length using the spatial three-dimensional positions of 95 areas and their connectivity. The total wiring length was between the case of only establishing the shortest-possible connections and establishing connections randomly regardless of distance (figure \ref{fig:wiring}{\it a}). A reduction of the wiring length was possible due to the number of long-distance connections in the original networks \citep{Kaiser2004c}; some of them even spanning almost the largest possible distance between areas. Why would these metabolically expensive connections exist in such large numbers? We tested the effect of removing all long-distance connections and replacing them by short-distance connections. Whereas several network measures improved, the value for the ASP increased when long-distance connections were unavailable (figure \ref{fig:wiring}{\it b}). Retaining a lower ASP has two benefits: First, there are fewer intermediate areas that might distort the signal. Second, as fewer areas are part of shortest paths, the transmission delay along a pathway is reduced. The propagation of signals over long distances, without any delay imposed by intermediate nodes, has an effect on synchronization as well: both nearby (directly connected) areas and faraway areas are able to get a signal at about the same time and could have synchronous processing \citep{Kaiser2006}. A low ASP might also be necessary because of the properties of neurons: John von Neumann, taking into account the low processing speed and accuracy of individual neurons, suggested that neural computation needed to be highly parallel with using a low number of subsequent processing steps \citep{vonNeumann1958}. But having a low ASP also brings a potential danger: How can it be prevented that information or activity flows uncontrolled through the entire network?

\subsection*{Balanced network activation through hierarchical connectivity}
Few processing steps enable the rapid transfer of activation patterns through cortical networks but this flow could potentially activate the whole brain. Such large-scale activations in the form of increased activity can be observed in the human brain during epileptic seizures: about 1\% of the population is currently affected by epilepsy. In contrast to computer networks with a continuous flow of viruses and spam e-mails, the brain has some built-in mechanisms for preventing large-scale activation.

An essential requirement for the representation of functional patterns in complex neural networks, such as the mammalian cerebral cortex, is the existence of stable network activations within a limited critical range. In this range, the activity of neural populations in the network persists between the extremes of quickly dying out, or activating a large part of the network as during epileptic seizures. The standard model would be to achieve such a balance by having interacting excitatory and inhibitory neurons. Whereas such models are of great value on the local level of neural systems, they are less meaningful when trying to understand the global level of connections between columns, areas, or area clusters.

Global corticocortical connectivity (connections between brain areas) in mammals possesses an intricate, nonrandom organization. Projections are arranged in clusters of cortical areas, which are closely linked among each other, but less frequently with areas in other clusters. Such structural clusters broadly agree with functional cortical subdivisions. This cluster organization is found at several levels: Neurons within a column, area or area cluster (e.g. visual cortex) are more frequently linked with each other than with neurons in the rest of the network \citep{Hilgetag2004}.

\begin{figure}
\includegraphics[width=\textwidth]{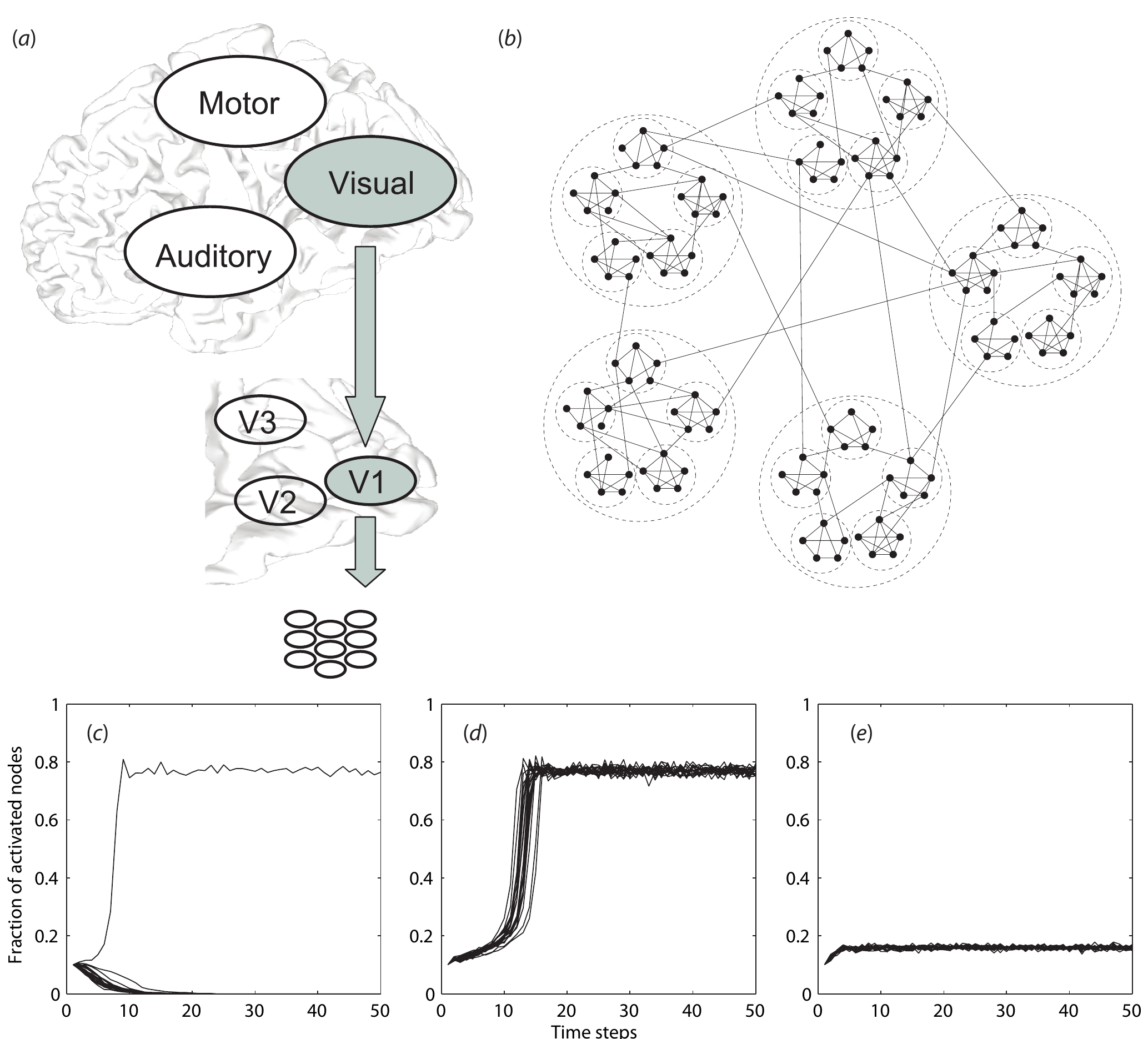}
\caption{({\it a}) The hierarchical network organization ranges from cluster such as the visual cortex to sub-cluster such as V1 to individual nodes being cortical columns. ({\it b}) Schematic view of a hierarchical cluster network with five clusters containing five sub-clusters each. Examples for spread of activity in ({\it c}) random, ({\it d}) small-world and ({\it e}) hierarchical cluster networks ($i=100, i_0=150$), based on 20 simulations for each network. \label{fig:spreading}}
\end{figure}

Using a basic spreading model without inhibition, we investigated how functional activations of nodes propagate through such a hierarchically clustered network \citep{Kaiser2007NJP}. The hierarchical network consisted of 1000 nodes made of 10 clusters with 100 nodes each. In addition, each cluster consisted of 10 sub-clusters with 10 nodes each (figure \ref{fig:spreading}{\it a, b}). Connections were arranged so that there were more links within (sub-)clusters than between (sub-)clusters. Starting with activating 10\% of randomly chosen nodes, nodes became activated if at least six directly connected nodes were active. Furthermore, at each time step, activated nodes could become inactive with a probability of 30\%.

The simulations demonstrated that persistent and scalable activation could be produced in clustered networks, but not in random or small-world networks of the same size (figure \ref{fig:spreading}{\it c-e}). Robust sustained activity also occurred when the number of consecutive activated states of a node was limited due to exhaustion. These findings were consistent for threshold models as well as integrate-and-fire models of nodes indicating that the topology rather than the activity model was responsible for balanced activity. In conclusion, hierarchical cluster architecture may provide the structural basis for the stable and diverse functional patterns observed in cortical networks. But how do networks with such properties arise?

\section{Design vs. Self-organization}
Neural systems, rather than being designed, evolved over millions of years. Starting from diffuse homogeneous networks, network clusters evolved when different tasks had to be implemented. During individual brain development, the architecture is formed by a combination of genetic blueprint and self-organization \citep{Striedter2005}.

What are the mechanisms of self-organization during network development? A possible algorithm for developing spatial networks with long-distance connections and small-world connectivity is spatial growth \citep{Kaiser2004b}. In this approach, the probability to establish a connection decays with the spatial (Euclidean) distance thereby establishing a preference for short-distance connections. This assumption is reasonable for neural networks as the concentration of growth factors decays with the distance to the source so that faraway neurons have a lower probability to detect the signal and sent a projection toward the source region of the growth factor. In addition, anatomical studies have shown that the probability of establishing a connections decreases with the distance between neurons.

In contrast to previous approaches that generated spatial graphs, the node positions were not determined before the start of connection establishment. Instead, starting with one node, a new node was added at each step at a randomly chosen spatial position. For all existing nodes, a connection between the new node $u$ and an existing node $v$ was established with probability
\begin{equation}\label{exponential}
P(u,v) = \beta \ e^{-\alpha \ d(u, v)},
\end{equation}
where $d(u, v)$ was the spatial distance between the node positions, and $\alpha$ and $\beta$ were scaling coefficients shaping the connection probability. A new node that did not manage to establish connections was removed from the network. Node generation was repeated until the desired number of nodes was established. Parameter $\beta$ ("density") served to adjust the general probability of edge formation. The nonnegative coefficient $\alpha$ ("spatial range") regulated the dependence of edge formation on the distance to existing nodes. Depending on the parameters $\alpha$ and $\beta$, spatial growth could yield networks similar to small-world cortical, scale-free highway-transportation networks as well as networks in non-Euclidean spaces such as metabolic networks \citep{Kaiser2004b}. Specifically, it was possible to generate networks with similar wiring organization than the macaque cortical network \citep{Kaiser2004c}. Using different time domains for connection development, where several spatial regions of the network establish connections in partly overlapping time windows, allows the generation of multiple clusters or communities \citep{Kaiser2007NC}.

\section{Outlook}
Natural neural systems, such as cortical networks of connections between brain regions, have developed several properties that are desirable for computers as well. Cortical networks show an innate ability to compensate for and recover from damages to the network. Whereas removing the few highly-connected nodes has a large effect on network structure, a random removal of nodes or edges has a small effect in most of the cases. In addition, the spatial layout of cortical and neuronal networks exhibiting several long-distance connections ensures few processing steps and thus a faster response time. Speculating about the future, these mechanism for robust and rapid processing might provide new ideas for artificial neural network as well as for computer architecture. As the 'programme' of the brain is implemented in its wiring organization, the topology of the brain might inspire theoretical work in the organization of parallel processing and integration.

Towards these topics, we currently work on three questions. First, to identify properties for robust processing in the brain. This includes understanding mechanisms for recovery in neural systems. These mechanisms will then be applied to computer networks to see if they can lead to faster recovery after failure.  Second, to investigate epileptic spreading in cortical networks. We intend to determine how the network structure influences activity or, for the disease state, seizure spreading in cortical networks. The more general analysis of spreading in networks could give useful insights in how to prevent virus spreading in communication networks. Finally, to find principles that guide the development of neural networks over time. By looking at general constraints for network development such as space, resources for connection establishment and maintenance, or global performance of a network, the reasons for normal and disturbed network development can be assessed. Ideally, this knowledge might lead to artificial neural networks with brain-like topology as well as processing.

In conclusion, we hope that future advances in our understanding of neural systems might (again) inspire solutions to problems in computer systems.

% REFERENCES
\bibliographystyle{myapalike}
\small

%\bibliographystyle{plain}
%\bibliography{c:/marcus/science/resources/neuro}
%\bibliography{neuro}

\newcommand{\noopsort}[1]{} \newcommand{\printfirst}[2]{#1}
  \newcommand{\singleletter}[1]{#1} \newcommand{\switchargs}[2]{#2#1}

\newpage
%[width=\textwidth]
\section*{Author Profile}

{\centering \includegraphics{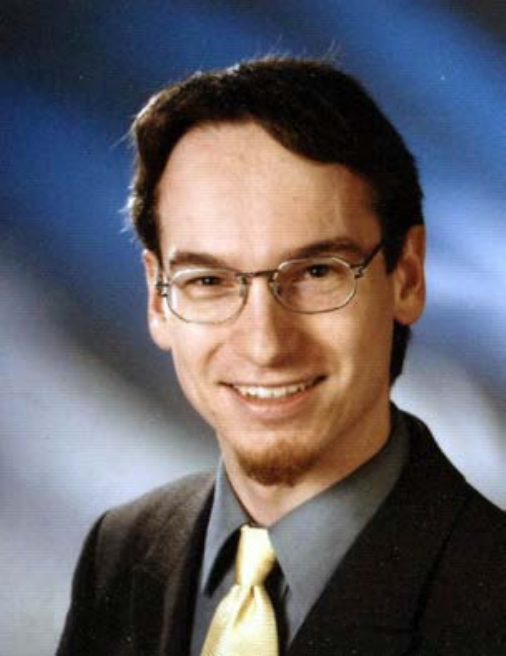} }

Born in Essen, Germany, Marcus Kaiser studied biology up to a Masters degree in 2002 at the Ruhr-University Bochum and continues studies of computer science at the distance university Hagen. He obtained his PhD in Neuroscience, funded by a fellowship from the German National Academic Foundation, from the Jacobs University Bremen in 2005. Directly after finishing his PhD, he started as RCUK Academic Fellow in Complex Neural Systems and Behaviour at Newcastle University. Having an appointment in Computing Science as well as at the Institute of Neuroscience, Marcus is interested in understanding the architecture and processing of the brain as a network. He works on development, spatial organisation, seizure spreading, and recovery in cortical networks. He gratefully acknowledges support from EPSRC (EP/E002331/1) and the Royal Society (RG/2006/R2). More information is available at http://www.biological-networks.org/

\end{document}